\RequirePackage{snapshot}
\documentclass[a4paper]{IEEEtran}

\usepackage[utf8]{inputenc}
\usepackage[british]{babel}
\usepackage{enumitem}
\usepackage{graphicx}
\usepackage{mathptmx}
\usepackage{amssymb}
\usepackage{microtype}
\usepackage{booktabs}
\usepackage{tabularx}
\usepackage[backend=biber,
  style=ieee,
  eprint=false,
  doi=false,
  isbn=false,
  url=false,
]{biblatex}
\usepackage{url}
\usepackage{hyperref}

\usepackage{xpatch}

\newcommand{\cL}{\mathcal{L}}
\newcommand{\cN}{\mathcal{N}}
\newcommand{\cT}{\mathcal{T}}
\newcommand{\set}[1]{\{#1\}}

\newcommand{\force}[2]{\mathit{Force}({#1},{#2})}
\newcommand{\ann}{\mathit{State}}

\AtBeginBibliography{\small}
\AtEveryBibitem{\clearfield{month}\clearfield{venue}}

\title{ICT Support for Regulatory Compliance\break of Business Processes}
\author{\IEEEauthorblockN{Guido Governatori}\\
\IEEEauthorblockA{NICTA, Australia\\
\url{guido.governatori@nicta.com.au}}
}
\date{}
\newlist{stmnt}{enumerate}{4}
\newenvironment{statement}
  {\begin{stmnt}[label=(S\arabic*),resume=stmnt,topsep=\abovedisplayskip]\item}
  {\end{stmnt}}
\newtheorem{definition}{Definition}
\newtheorem{example}{Example}

\newcommand{\PERM}{\mathrm{P}}
\newcommand{\OM}{\mathrm{OM}}
\newcommand{\OAPP}{\mathrm{OAPP}}
\newcommand{\OAPNP}{\mathrm{OAPNP}}
\newcommand{\OANPP}{\mathrm{OANPP}}
\newcommand{\OANPNP}{\mathrm{OANPNP}}

\begin{document}

\maketitle\thispagestyle{empty}
 \begin{abstract}
  In this paper we propose an ITC (Information and Communication Technology)
  approach to support regulatory compliance for business processes, and we
  report on the development and evaluation of a business
  process compliance checker called Regorous, based on the
  compliance-by-design methodology
  proposed by Governatori and Sadiq \cite{HandbookBPM:IGI}.
\end{abstract}

\section{Introduction} 
\label{sec:introduction}

Regulatory compliance is the set of activities an enterprise does to ensure
that its core business does not violate relevant regulations, in the
jurisdictions in which the business is situated, governing the (industry)
sectors where the enterprise operates.

The activities an organisation does to achieve its business objectives can be
understood as business processes, and consequently they can be represented by
business process models. On the other hand a normative document (e.g., a code,
a bill, an act) can be understood as a set of clauses, and these clauses can
be represented in an appropriate formal language. Based on this \cite{edoc06}
proposed that \emph{business process compliance} is a relationship between the
formal representation of a process model and the formal representation of a
relevant regulation. The specific relationship is that the formal specifications
for the business process do not violate the conditions set out by the 
formal specifications modelling the regulation.  

To gain compliance different strategies can be devised. \cite{Handbook}
classifies approaches to compliance as \emph{detective}, \emph{corrective} and
\emph{preventative}.

\emph{Detective measures} are intended to identify ``after-the-fact''
un-compliant situations. There are two main approaches: (a) \emph{retrospective
reporting} through manual audits by consultants or through IT forensics and
Business Intelligence tools; (b) \emph{automated detections} generating audit
reports against hard-coded checks performed on the requisite system. Unlike
the first approach, automated detection reduces the assessment time and
consequently also the time of un-compliance remediation/mitigation.

\emph{Corrective measures} are intended to limit the extent of any consequence
caused by un-compliant situations. For example, situations that can arise from
the introduction of a new norm impacting upon the business, to the
organisation coming under surveillance and scrutiny by a control authority or
to an enforceable undertaking.

The two approaches above suffer from lack of \emph{sustainability}, caused by
the extreme interest of companies in continuous improvements of the quality of
services, and for changing legislations and compliance requirements. Indeed,
even with automated detection means, the hard coded checking of repositories can
quickly grow to a very large scale making it extremely difficult to evolve and
maintain. To obviate these problem \cite{bpm07,LuSadGov:bpd07} propose a
\emph{preventative focus} based on the idea of \emph{compliance-by-design}.

The key aspect of the compliance-by-design methodology is to supplement
business process models with additional information to ensure that a business
process is compliant with relevant normative frameworks before the deployment
of the process itself.

From the previous discussion it should be clear that for an effective and 
successfully application of ICT (Information and Communication Technology) 
techniques to the problem of ensuring that 
business processes are compliant with the relevant regulations we need two 
components: (i) a conceptually sound formal representation of a business 
process and (ii) a conceptually sound formalism to model and to reasoning
with norms.   In Section \ref{sec:business_process_modelling} we will recall 
the basic of business process modelling. In Section \ref{sec:norms} we propose
a model of norms which provides a conceptually sound, rich and comprehensive 
classification of normative concepts (i.e., obligations, prohibitions, 
permission and violation) described in terms of processes. Each notion 
introduced in this section is justified by a concrete case taken from 
existing statutory acts, regulations or other legally binding documents.
Section~\ref{sec:compliance} is dedicated to give proper definitions of what it
means for a process to be compliant with a given set of norms. 
In Section~\ref{sec:bpcc_architecture} we describe the architecture of Regorous 
Process Designer, a compliance checker based on the methodology proposed by 
Governatori and Sadiq \cite{HandbookBPM:IGI}. Section~\ref{sec:evaluation} 
describes the implementation of Regorous and reports on an industry scale case
study which has been used to empirically evaluate the approach. We conclude the 
paper with a short discussion of how the proposed approach can be used in
different phases of the lifecycle of a process and relationships with 
monitoring and auditing (Section \ref{sec:extensions}). Section
\ref{sec:conclusions} quickly discusses some closely related work.


\section{Business Process Modelling} 
\label{sec:business_process_modelling}

In this section we provide the vary basics of business process modelling, for an
extensive presentation see \cite{bpm-textbook}. 
A business process model is a self-contained, temporal and logical order in
which a set of activities are executed to achieve a business goal. Typically a
process model describes what needs to be done and when (control flow), who is
going to do what (resources), and on what it is working on (data). Many
different formalisms (Petri-Nets, Process algebras, \dots) and notations
(BPMN, YAWL, EPC, \dots) have been proposed to represent business process
models. Besides the difference in notation, purposes, and expressive power,
business process languages typically contain the following minimal set of
elements:
\begin{itemize}
  \item tasks
  \item connectors
\end{itemize}  
where a task corresponds to a (complex) business activity, and connectors
(e.g., sequence, and-join, and-split, (x)or-join, (x)or-split) define the
relationships among tasks to be executed. The combination of tasks and
connectors defines the possible ways in which a process can be executed. Where
a possible execution, called \emph{process trace} or simply \emph{trace}, is a
sequence of tasks respecting the order given by the connectors.

\begin{figure}[htb]
    \includegraphics[width=\columnwidth]{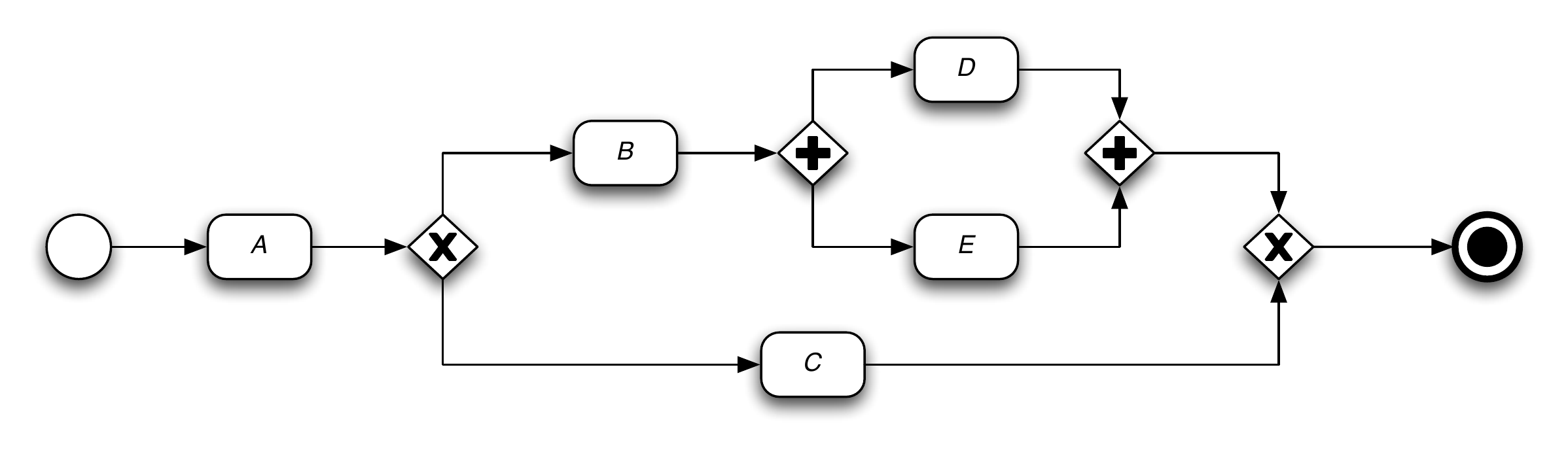}
  \caption{Example of a business process model in standard BPMN notation}
  \label{fig:process}
\end{figure}  
Consider the process in Figure~\ref{fig:process}, in standard BPMN notation,
where we have a task $A$ followed by an xor split. In the xor split in one of
the branches we have task $B$ followed by the and-split of a branch with task
$D$, and a brach consisting of only task $E$. The second branch of the
xor-split has only one task: $C$. The traces corresponding to the process are
$\langle A,C\rangle$, $\langle A, B, D, E\rangle$ and $\langle A,B, E,
D\rangle$. Given a process $P$ we will use $\cT_{P}=\set{t_{1}, t_{2}, \dots}$
to denote the set of traces of $P$. 

Compliance is not only about the
tasks an organisation has to perform to achieve its business goals, but
it is also concerned on their effects (i.e., how the activities in the tasks
change the environment in which they operate), and the artefacts produced by
the tasks (for example, the data resulting from executing a task or
modified by the task) \cite{ruleml2012}. To capture this aspect
\cite{bpm07} proposed to enrich process models with semantic annotations. Each
task in a process model can have attached to it a set of semantic annotations.
An annotation is just a set of formulas giving a (partial) description of the
environment in which a process operates. Then, it is possible to associate to
each task in a trace a set of formulas corresponding to the state of the
environment after the task has been executed in the particular trace. Notice,
that different traces can results in different states, even if the tasks in the
traces are the same. In addition, even if the end states are the same, the
intermediate states can be different. Accordingly, we extend the notion of
trace. First of all, we introduce the function
\[
\ann\colon \cT_{P}\times \mathbb{N}\mapsto 2^{\cL},
\] 
where $\cL$ is the set of formulas of the language used to model the
annotations. Let us illustrate with an example the meaning of the function
$\ann$. Suppose we have the trace $t=\langle A, B, D, E\rangle$, and
that $\ann(t,3)=\set{p,q,r}$. This means that $\set{p,q,r}$ is the
state resulting after executing $D$ in the trace $t$ ($D$ is the third task in
$t$). Notice that a trace uniquely determines the sequence of states obtained
by executing the trace. Thus, in what follows we use a trace to refer to a
sequence of tasks, and the corresponding sequence of states.

\section{Normative Requirements}
\label{sec:norms}

\subsection{Modelling Norms}
\label{sec:modelling_norms}
The scope of norms is to regulate the behaviour of their subjects and to define
what is legal and what is illegal. Norms typically describe the conditions
under which they are applicable and the normative effects they produce when
applied. A comprehensive list of normative effects is provided in
\cite{ruleml2009}. In a compliance perspective, the
normative effects of importance are the deontic effects (also called normative
positions). The basic deontic
effects are: \emph{obligation}, \emph{prohibition} and
\emph{permission}.\footnote{There are other deontic effects, but these can be
derived from the basic ones, see \cite{Sartor05}.}

Let us start by consider the basic definitions for such
concepts:\footnote{Here we consider the definition of such concepts given by
the OASIS LegalRuleML working group. The OASIS LegalRuleML glossary is
available at
\url{http://www.oasis-open.org/apps/org/workgroup/legalruleml/download.php/48435/Glossary.doc}.}
\begin{description}
  \item[Obligation] A situation, an act, or a course of action to which a
    bearer is legally bound, and if it is not achieved or performed
    results in a violation.
  \item[Prohibition] A situation, an act, or a course of action which a
    bearer should avoid, and if it is achieved results in a violation.
  \item[Permission] Something is permitted if the obligation or the
    prohibition to the contrary does not hold.
\end{description}

Obligations and prohibitions are constraints that limit the space of action of
processes; the difference from other types of constraints is that they can
be violated, and a violation does not imply an inconsistency within a process
with the consequent termination of or impossibility to continue the business
process. Furthermore, it is common that
violations can be compensated for, and processes with compensated violations
are still compliant \cite{GM05,HandbookBPM:IGI,ism08}; for example contracts typically contain
compensatory clauses specifying penalties and other sanctions triggered by
breaches of other contract clauses \cite{coala}. Not all violations are
compensable, and the presence of uncompensated violations means that a process is not
compliant. Permissions cannot be violated, thus permissions do not play a
direct role in compliance; they can be used to determine that there are no
obligations or prohibitions to the contrary, or to derive other deontic effects.
Legal reasoning and legal theory typically assume a strong relationship
between obligations and prohibitions: the prohibition of $A$ is the
obligation of $\neg A$ (the opposite of $A$), and then if $A$ is obligatory,
then $\neg A$ is forbidden \cite{Sartor05}. In this paper we will subscribe to
this position, given that our focus here is not on how to determine
what is prescribed by a set of norms and how to derive it. Accordingly, we
can restrict our analysis to the notion of \textit{obligation}.

Compliance means to identify whether a process violates or not a set of
obligations. Thus, the first step is to determine whether and when an
obligation is in force. Hence, an important aspect of the
study of obligations is to understand the lifespan of an obligation and the
consequence it has on the activities carried out in a process. As we have alluded to
above norms give the conditions of applicability of obligations. The question
then is how long does an obligation hold for, and based on this there are
different conditions to fulfill the obligation. We take a systematic approach
to this issue. A norm can specify that an obligation is in force for a
particular time point or, more often, a norm indicates when an obligation
enters in force. An obligation remains in force until terminated or removed.
Accordingly, in the first case we will speak of \emph{punctual obligations}
and in the second case of \emph{persistent obligations}.   

For persistent obligations we can ask if to fulfill an obligation we have to
obey to it for all instants in the interval in which it is in force,
\emph{maintenance obligations}, or whether doing or achieving the content of
the obligation at least once is enough to fulfill it, \emph{achievement obligations}.
For achievement obligations another aspect to consider is whether the obligation
could be fulfilled even before the obligation is actually in force. If this is
admitted, then we have a \emph{preemptive obligation}, otherwise the
obligation is \emph{non-preemptive}.

The final aspect we want to touch upon in this section is the
termination of obligations. Norms can specify the interval in which an
obligation is in force. Previously, we discussed that what differentiates
obligations and other constraints is that obligations can be violated. What are
the effects of a violation on the obligation the violation violates? More
precisely, does a violation terminate the violated obligation? Meaning, do we
still have to comply with a violated obligation? If we do --the obligation
persists after being violated-- we speak of a \emph{perdurant obligation}, if it
does not, then we have a \emph{non-perdurant obligation}.

The classification discussed above is exhaustive. It has been obtained in a
systematic and comprehensive way
when one considers the aspect of the validity of obligations --or
prohibitions-- (i.e., whether they persist after they enter in force or they
are valid only for a specific time unit),
and the effects of violations on them, namely: whether a violation can be
compensated for, and whether an obligation persists after being violated. In the
next section we will provide formal definitions for the notions
introduced in this section and for each case we will show examples
taken form statutory acts and other legally binding documents.

\subsection{Modelling Obligations} 
\label{sec:modelling_obligations}

In this section we provide the formal definitions underpinning the
notion of compliance. In particular we formally define the different types 
of obligations introduced in Section~\ref{sec:modelling_norms}. 

\begin{definition}[Obligation in force]\label{def:force}
   Given a process $P$, and a trace $t\in\cT_P$. We define a function
   \[
     \mathit{Force}\colon\cT_P\times \mathbb{N}\mapsto 2^{\cL}.
   \]
\end{definition}

The function $\mathit{Force}$ associates to each task in a trace a set of
literals, where these literals represent the obligations in force for that
combination of task and trace. These are among the obligations that the
process has to fulfill to comply with a given normative framework. In the rest
of the section we are going to give definitions specifying when the process has
to fulfill the various obligations (depending on their type) to be deemed
compliant.

\begin{definition}[Punctual Obligation]\label{def:punctual}
  Given a process $P$ and a trace $t\in\cT_P$,
  an obligation $o$ is a \emph{punctual obligation} in $t$ if and only if
  $\exists n\in\mathbb{N}$ such that
  \begin{enumerate}
    \item $o\notin\force{t}{n-1}$,
    \item $o\notin\force{t}{n+1}$, and
    \item $o\in\force{t}{n}$.
  \end{enumerate}
  A punctual obligation $o$ is \emph{violated} in $t$ if and only if 
  $o\notin\ann(t,n)$.\footnote{For the conditions defining when an obligation 
  is violated we assume the same conditions defining the type of the
  obligation. For example, in this case $\exists n\in\mathbb{N}$ such that
  $o\in\force{t}{n}$.}
\end{definition}

A punctual obligation is an obligation that is in force in one task of a trace
(it might be the case that there are multiple instances in which the
obligation is in force). The obligation is violated if what the obligation
prescribes is not achieved in or done by the task, where this is represented by
the literal not being in the set of literals associated to the task in the
trace.

\begin{definition}[Achievement Obligation]\label{def:achievemnt}
  Given a process $P$ and a trace $t\in\cT_P$,
  an obligation $o$ is an \emph{achievement obligation} in $t$ if and only if
  $\exists n,m\in\mathbb{N}, n<m$ such that 
  \begin{enumerate} 
    \item $o\notin\force{t}{n-1}$,
    \item $o\notin\force{t}{m+1}$, and
    \item $\forall k: n\leq k\leq m, o\in\force{t}{k}$
  \end{enumerate}
  An achievement obligation $o$ is \emph{violated} in $t$ if and only if 
  \begin{itemize}[topsep=0pt]
    \item $o$ is preemptive and $\forall k\colon k\leq m,\ o\notin\ann(t,k)$;
    \item $o$ is non-preemptive and $\forall k\colon n\leq k\leq m,\
      o\notin\ann(t,k)$.
  \end{itemize}
\end{definition}

An achievement obligation is in force in a contiguous set of tasks in a trace.
The violation depends on whether we have a preemptive or a non-preemptive
obligation. A preemptive obligation $o$ is violated if no state
before the last task in which $o$ is in force has $o$ in its annotations;
for a non-preemptive obligation the set of states is restricted to those
defined by the interval in which the obligation is in force.

\begin{example}\label{ex:punctual}
Australian Telecommunications Consumers Protection Code 2012 (TCPC 2012). Article 8.2.1.\\
A Supplier must take the following actions to enable this outcome:
\begin{itemize}
  \item[(a)] \textbf{Demonstrate fairness, courtesy, objectivity and
    efficiency:} Suppliers must demonstrate, fairness and courtesy,
    objectivity, and efficiency by:
  \begin{itemize}
    \item[(i)]  Acknowledging a Complaint:
    \begin{itemize}
      \item[A.] immediately where the Complaint is made in person or
        by telephone;
      \item[B.] within 2 Working Days of receipt where the Complaint
        is made by email; \dots.
    \end{itemize}
  \end{itemize}
\end{itemize}  
The obligation to acknowledge a compliant made in person or by phone
(8.2.1.a.i.A) is a punctual obligation, since it has to be done `immediately'
while receiving it (thus it can be one of the activities done in the task
`receive complaint'). 8.2.1.a.i.B on the other hand is an achievement
obligation since the clause gives a deadline to achieve it. In addition it is
a non-preemptive obligation. It is not possible to acknowledge a complaint
before having it.
\end{example}

\begin{example}\label{ex:preemptive}
  Anti-Money Laundering and Counter-Terrorism Financing Act 2006. Clause 54
  (Timing of reports about physical currency movements).
  \begin{itemize}
    \item[(1)] A report under section 53 must be given:
    \begin{itemize}
      \item[(a)] if the movement of the physical currency is to be
        effected by a person bringing the physical currency into
        Australia with the person---at the time worked out under
        subsection (2); or
      \item[{[\dots\unkern]}]
      \item[(d)] in any other case---at any time before the movement
        of the physical currency takes place.
    \end{itemize}
  \end{itemize}
Clause (d) illustrates a preemptive obligation. The obligation is in force
when a financial transaction occurs, and the clause explicitly requires the
report to be submitted to the relevant authority \emph{before} the transaction
actually occurs (it might be the case that the transaction never occurs).
\end{example}

\begin{definition}[Maintenance Obligation]\label{def:maintenance}
  Given a process $P$ and a trace $t\in\cT_P$,
  an obligation $o$ is a \emph{maintenance obligation} in $t$ if and only if
  $\exists n,m\in\mathbb{N}$, $n<m$ such that:
  \begin{enumerate}
    \item $o\notin\force{t}{n-1}$,
    \item $o\notin\force{t}{m+1}$, and
    \item $\forall k\colon n\leq k\leq m, o\in\force{t}{k}$
  \end{enumerate}
  A maintenance obligation $o$ is \emph{violated} in $t$ if and only if 
  \[
    \exists k\colon n\leq k\leq m, o\notin\ann(t,k).
  \]
\end{definition}

Similarly to an achievement obligation, a maintenance obligation is in force
in an interval. The difference is that the obligation has to be complied with
for all tasks in the interval, otherwise we have a violation.

\begin{example}\label{ex:maintenance}
TCPC 2012. Article 8.2.1.\\
A Supplier must take the following actions to enable this outcome:
\begin{itemize}
  \item[(v)] not taking Credit Management action in relation to a specified
    disputed amount that is the subject of an unresolved Complaint in
    circumstances where the Supplier is aware that the Complaint has not
    been Resolved to the satisfaction of the Consumer and is being
    investigated by the Supplier, the TIO or a relevant recognised third
    party;
\end{itemize}
In this example, as it is often the case, a maintenance obligation implements
a prohibition. Specifically, it describes the prohibition to initiate a
particular type of activity until either a particular event takes place or a
state is reached.
\end{example}

The next three definitions are meant to capture the notion of compensation of
a violation. The idea is that a compensation is a set of penalties or
sanctions imposed on the violator, and fulfilling them makes amend for the
violation. The first step is to define what a compensation is. A compensation
is a set of obligations in force after a violation of an obligation
(Definitions~\ref{def:compensation} and \ref{def:compensable}). Since the
compensations are obligations themselves they can be violated, and they can be
compensable as well, thus we need a recursive definition for the notion of
compensated obligation (Definition~\ref{def:compensated}).
\begin{definition}[Compensation]\label{def:compensation}
  A \emph{compensation} is a function $\mathit{Comp}\colon \cL\mapsto 2^{\cL}$.
\end{definition}

\begin{definition}[Compensable Obligation]\label{def:compensable}
  Given a process $P$ and a trace $t\in\cT_P$,
  an obligation $o$ is \emph{compensable} in $t$ if and only if 
  \begin{enumerate}
    \item $\mathit{Comp}(o)\neq\emptyset$ and
    \item $\forall o'\in\mathit{Comp}(o), \exists n\in\mathbb{N}\colon o'\in\force{t}{n}$.
  \end{enumerate}
\end{definition}

\begin{definition}[Compensated Obligation]\label{def:compensated}
  Given a process $P$ and a trace $t\in\cT_P$, an obligation $o$ is
  \emph{compensated} in $t$ if and only if it is violated and for every
  $o'\in\mathit{Comp}(o)$ either:
  \begin{enumerate}
    \item $o'$ is not violated in $t$, or
    \item $o'$ is compensated in $t$.
  \end{enumerate}
\end{definition}

For a stricter notion, i.e., a compensated compensation does not amend the
violation the compensation was meant to compensate, we can simply remove the
recursive call, thus removing clause 2 from the above condition.

Compensations can be used for two purposes. The first is to specify
alternative, less ideal, outcomes. The second is to capture sanctions and
penalties. Examples~\ref{ex:compensation} and \ref{ex:yawl} below illustrate,
respectively, these two usages.

\begin{example}\label{ex:compensation}
 TCPC 2012. Article 8.1.1.\\
A Supplier must take the following actions to enable this outcome:
\begin{itemize}
  \item[(a)] \textbf{Implement a process}: implement, operate and comply
    with a Complaint handling process that:
  \begin{itemize}
    \item[(vii)] requires all Complaints to be:
    \begin{itemize}
      \item[A.] Resolved in an objective, efficient and fair manner; and
      \item[B.] escalated and managed under the Supplier's internal
        escalation process if requested by the Consumer or a former
        Customer.
    \end{itemize} 
  \end{itemize}
\end{itemize} 
\end{example}

\begin{example}\label{ex:yawl}
  YAWL Deed of Assignment, Clause 5.2.\footnote{\url{http://www.yawlfoundation.org/files/YAWLDeedOfAssignmentTemplate.pdf}, retrieved on March 28, 2013.}\\
  Each Contributor indemnifies and will defend the Foundation against any
  claim, liability, loss, damages, cost and expenses suffered or incurred by
  the Foundation as a result of any breach of the warranties given by the
  Contributor under \textbf{clause 5.1}.
\end{example}

The final definition is that of perdurant obligation. The intuition behind it
is that there is a deadline by when the obligation has to be fulfilled. If it
is not fulfilled by the deadline then a violation is raised, but the
obligation is still in force. Typically, the violation of a perdurant
obligation triggers a penalty, thus if the perdurant obligation is not
fulfilled in time, then the process has to account for the original
obligation as well as the penalties associated with the violation.

\begin{definition}[Perdurant Obligation]\label{def:perdurant}
  Given a process $P$ and a trace $t\in\cT_P$,
  an obligation $o$ is a \emph{perdurant obligation} in $t$ if and only if
  $\exists n,m\in\mathbb{N}$, $n<m$ such that 
  \begin{enumerate} 
    \item $o\notin\force{t}{n-1}$,
    \item $o\notin\force{t}{m+1}$, and 
    \item $\forall k\colon n\leq k\leq m, o\in\force{t}{k}$.
  \end{enumerate}
  A perdurant obligation $o$ is \emph{violated} in $t$ if and only if 
  \[
    \exists k\colon n< k<m,\ \forall j, j\leq k,\  o\notin\ann(t,j)
  \]
\end{definition}

Consider again Example~\ref{ex:punctual}. Clauses TCPC 8.2.1.a.i.A and
8.2.1.a.i.B state what are the deadlines to acknowledge a complaint, but
8.2.1.a.i prescribes that complaints have to be acknowledged. Thus, if a
complaint is not acknowledged within the prescribed time then either clause A
or B are violated, but the supplier still has the obligation to acknowledge
the complaint. Thus the obligation in clause (i) is a perdurant obligation.

\section{Modelling Compliance}
\label{sec:compliance}

The set of traces of a given business process describes the behavior of the
process insofar as it provides a description of all possible ways in
which the process can be correctly executed. Accordingly, for the purpose of
defining what it means for a process to be compliant, we will consider a
process as the set of its traces.

Intuitively a process is compliant with a normative system\footnote{Here, by
normative system we simply mean a set of norms, where a norm is a formula in
the underlying (deontic) language. For a business process the normative system
could vary from a particular regulation, to a specific statutory act, a set of
best practices, a standard, simply a policy internal to an organisation or a
combination of these types of prescriptive documents.} if
it does not breach the normative system. Given that, in
general, it is possible to perform a business process in many different ways,
thus we can have two notions of compliance, namely:
\begin{statement}\label{stm:full}
  A process is (fully) compliant with a normative system if it is impossible to
  violate the normative system while executing the process.
\end{statement}
The intuition about the above condition is that no matter in which way the
process is executed, its execution does not violate the normative
system. For the second one we consider the case that there is an execution of
the process that does not violate the norms.
\begin{statement}\label{stm:partial}
  A process is (partially) compliant with a normative system if it is
  possible to execute the process without violating the normative system.
\end{statement}
Based on the above intuition we can give the following definition:
\begin{definition}\label{def:compliance}
  Let $\cN$ be a normative system. 
  \begin{enumerate}
    \item A process $P$ \emph{fully complies} with $\cN$ if and only if every 
      trace $t\in\cT_{P}$ complies with $\cN$.
    \item A process $P$ \emph{partially complies} with $\cN$ if and only if 
      there is a trace $t\in\cT_{P}$ that complies with $\cN$.
  \end{enumerate}
\end{definition}
Notice that in \ref{stm:full} and \ref{stm:partial} compliance means ``lack of
violations'' while in Definition \ref{def:compliance} we had ``comply with''.
For the purpose of this paper we will treat these two concepts as equivalent.
More precisely
they are related by the following definition.

\begin{definition}\label{def:trace-compliance}
  A trace $t$ complies with a normative system $\cN=\set{n_{1}, n_{2},\dots}$
  if and only if all norms in $\cN$ have not been violated.
\end{definition}

In Section~\ref{sec:modelling_norms} we introduced various types
of norms and for each type we described its semantics in terms of 
what constitutes a violation of a norm of that type.

The possibility of a norm to be violated is what distinguish norms from other
types of constraints. Then, given that violations are possible, one has to
consider that violations can be compensated. Is a process where some norms have
violated and compensated for compliant? To account for this possibility we
introduce the distinction between \emph{strong} and \emph{weak} compliance.
Strong compliance corresponds to Definition \ref{def:trace-compliance}.
Weak compliance is defined as follows:

\begin{definition}\label{def:weak}
  A trace $t$ is weakly compliant with a normative system $\cN$ if and only if
  every violated norm has been compensated for.
\end{definition}

\section{Regorous Architecture} 
\label{sec:bpcc_architecture}

In this section we introduce the architecture of Regorous Process
Designer (from now on simply Regorous), a business
process compliance checker based on the
methodology proposed by Governatori and Sadiq \cite{HandbookBPM:IGI}.

As we have already discussed to check whether a business process is compliant
with a relevant regulation, we need an annotated business process model and
the formal representation of the regulation. The annotations are attached to
the tasks of the process, and they can be used to record the data, resources and
other information related to the single tasks in a process.

For the formal representation of the regulation we use FCL
\cite{coala,apccm2010}. FCL is a simple, efficient,
flexible rule based logic. FCL has been obtained from the combination of
defeasible logic (for the efficient and natural treatment of exceptions, which
are a common feature in normative reasoning) \cite{tocl} and a deontic logic of
violations \cite{violation}. In FCL norms are represented by rules with the 
following form
\[
  a_{1},\dots,a_{n} \Rightarrow c
\]  
Where $a_{1},\dots,a_{n}$ are the conditions of applicability of the norm/rule
and $c$ is the \emph{normative effect} of the norm/rule. FCL distinguishes two
normative effects: the first is that of introducing a definition for a new
term. For example the rule
\[
  \mathit{customer}(x), \mathit{spending}(x)>1000
  \Rightarrow\mathit{premium\_customer}(x)
\]
specifies that, typically, a premium customer is a customer who has spent over
1000 dollars. The second normative effect is that of triggering obligations and
other deontic notions.  FCL supports all deontic notions presented in 
Section~\ref{sec:modelling_norms}, in addition it has mechanisms to terminate 
and remove obligations (see \cite{apccm2010} for full details). For
obligations and permission we use the following notation:
\begin{itemize}
\item $[\PERM]p$: $p$ is permitted;
\item $[\OM]p$: there is a maintenance obligation for $p$;
\item $[\OAPP]p$: there is an achievement preemptive and perdurant obligation for $p$;
\item $[\OAPNP]p$: there is an achievement preemptive and non-perdurant obligation for $p$;
\item $[\OANPP]p$: there is an achievement non preemptive and perdurant obligation for $p$;
\item $[\OANPNP]p$: there is an achievement non preemptive and non-perdurant obligation for $p$.
\end{itemize}

Compensations are implemented based on the notion of `reparation chain'
\cite{violation}. A
reparation chair is an expression $O_{1}c_{1}\otimes O_{2}c\otimes
\cdots\otimes O_{n}c_{n}$, where each $O_{i}$ is an obligation, and each
$c_{i}$ is the content of the obligation (modelled by a literal). The meaning
of a reparation chain is that we have that $c_{1}$ is obligatory, but if the
obligation of $c_{1}$ is violated, i.e., we have $\neg c_{1}$, then the
violation is compensated by $c_{2}$ (which is then obligatory). But if even
$O_{2}c_{2}$ is violated, then this violation is compensated by $c_{3}$ which,
after the violation of $c_{2}$, becomes obligatory, and so on.

It is worth noticing that FCL allows deontic expressions (but not reparation
chains) to appear in the body of rules, thus we can have rules like:
\[
  \mathit{restaurant}, [\PERM]\mathit{sell\_alcohol} \Rightarrow 
    \begin{array}[t]{@{}l}
      {}[\OM]\mathit{show\_license} \otimes {}\\ 
      {}[\OAPNP] \mathit{pay\_fine}.
    \end{array}
\]
The rule above means that if a restaurant has a license to sell alcohol (i.e.,
it is permitted to sell it, $[\PERM]\mathit{sell\_alcohol}$), then it has a
maintenance obligation to expose the license ($[\OM] \mathit{show\_license}$),
if it does not then it has to pay a fine ($[\OAPNP] \mathit{pay\_fine}$). The
obligation to pay the fine is non-pre-emptive (this means it cannot be paid
before the violation). For full description of FCL and its feature see
\cite{coala,apccm2010}.

Finally, FCL is agnostic about the nature of the literals it uses. They can
represent tasks (activities executed in a process) or propositions representing
state variables.

Compliance is not just about the tasks to be executed in a process but also on
what the tasks do, the way they change the data and the state of artefacts
related to the process, and the resources linked to the process. Accordingly,
process models must be enriched with such information. \cite{bpm07} proposes
to enrich process models with semantic annotations. Each task in a process
model can have attached to it a set of semantic annotations. In our approach
the semantic annotations are literals in the language of FCL, representing the
effects of the tasks. The approach can be used to model business process data
compliance \cite{ruleml2012}.

\begin{figure}[htbp]
  \centering
    \includegraphics[width=.98\linewidth]{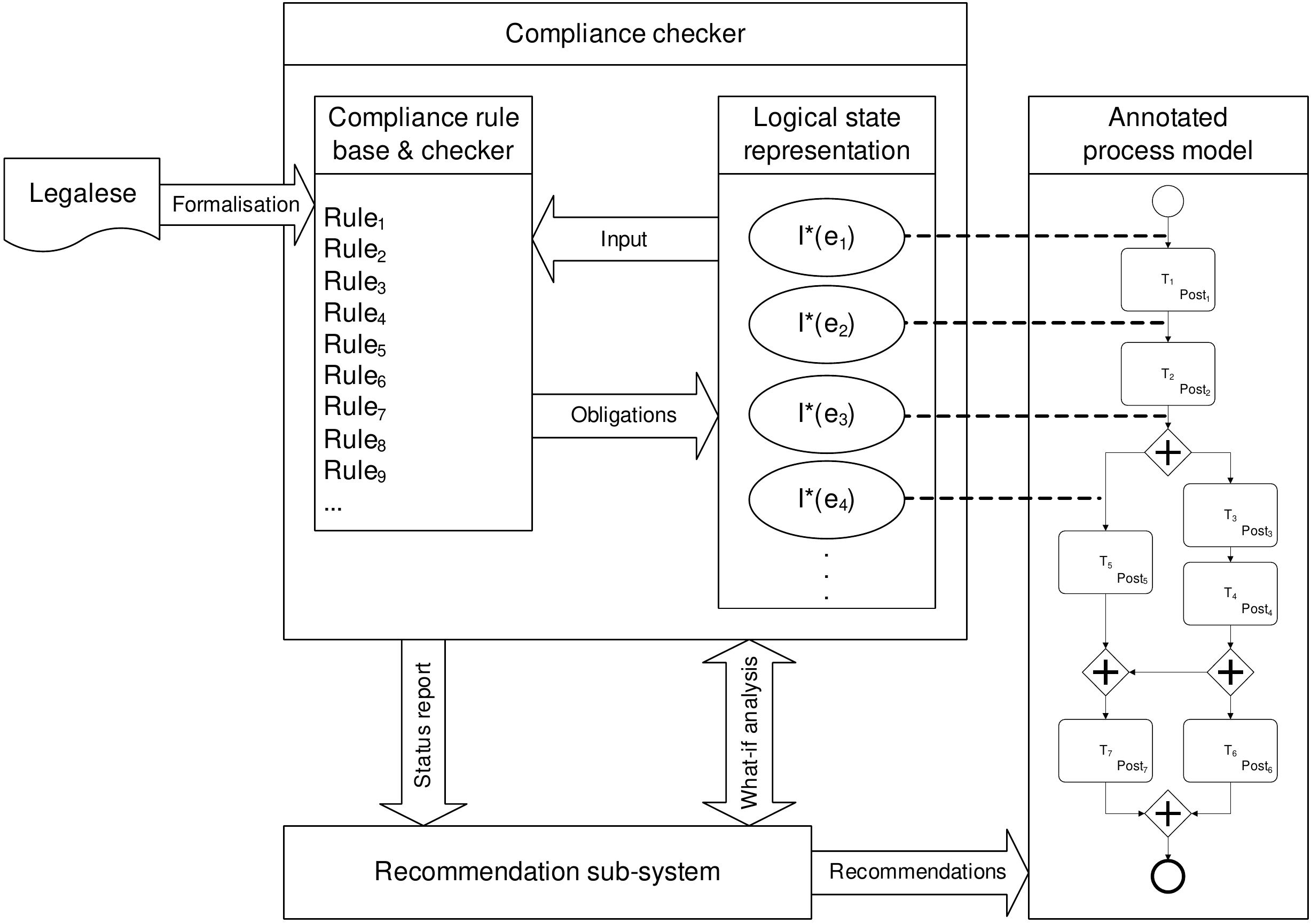}
  \caption{Architecture of Regorous}
  \label{fig:Figures_Overview_picture}
\end{figure}

Figure~\ref{fig:Figures_Overview_picture} depicts the architecture of Regorous.
Given an annotated process and the formalisation of the relevant
regulation, we can use the algorithm propose in \cite{algorithm,apccm2010} to
determine whether an annotated process model is compliant. The process runs
as follows:
\begin{itemize}
  \item Generate an execution trace of the process.
  \item Traverse the trace:
  \begin{itemize}
    \item for each task in the trace, cumulate the effects of the
      task using an update semantics (i.e., if an effect in
      the current task conflicts with previous annotations, update using
      the effects of the current tasks).
    \item use the set of cumulated effects to determine which
      obligations enter into force at the current tasks. This is done by
      a call to an FCL reasoner. 
    \item add the obligations obtained from the previous step to the set
      of obligations carried over from the previous task.
    \item determine which obligations have been fulfilled, violated, or
      are pending; and if there are violated obligations check whether
      they have been compensated.
  \end{itemize}
  \item repeat for all traces.
\end{itemize}
A process is compliant if and only if all traces are compliant (all
obligations have been fulfilled or if violated they have been compensated). A
process is weakly compliant if there is at least one trace that is compliant.

\begin{figure*}[htbp]
  \centering
    \includegraphics[width=\textwidth]{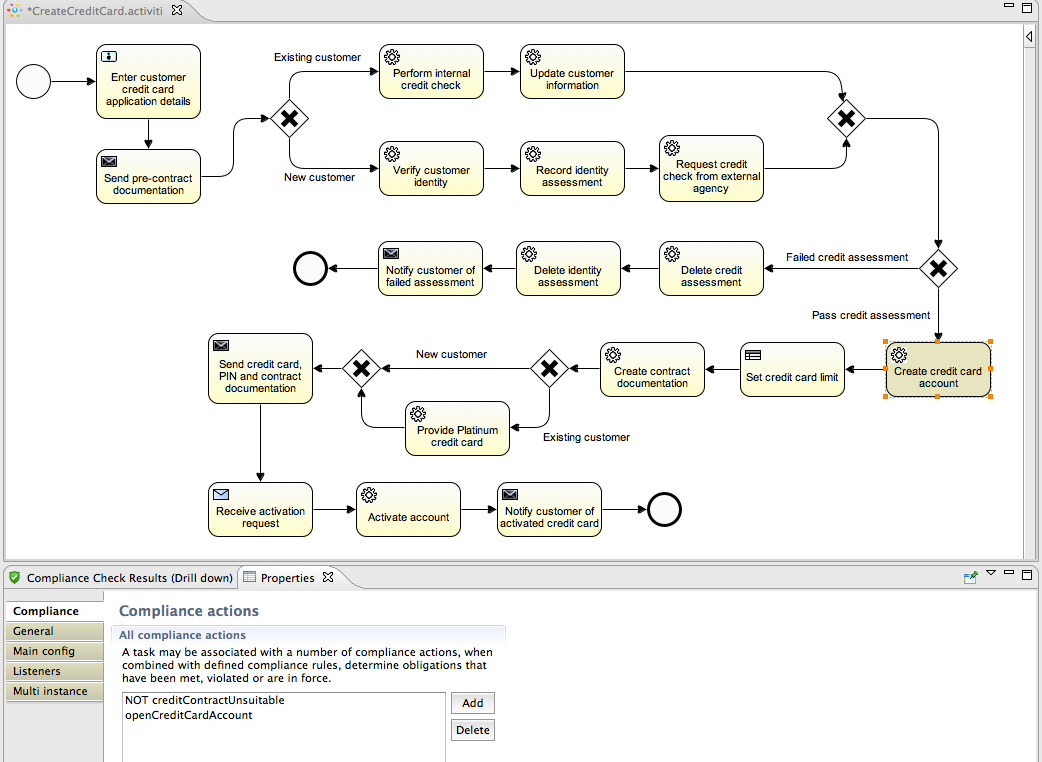}
  \caption{An Opening Credit Card Account Process with Annotations in Regorous}
  \label{fig:Figures_bpcc-general}
\end{figure*}

\begin{figure*}[htbp]
  \centering \includegraphics[width=\textwidth]{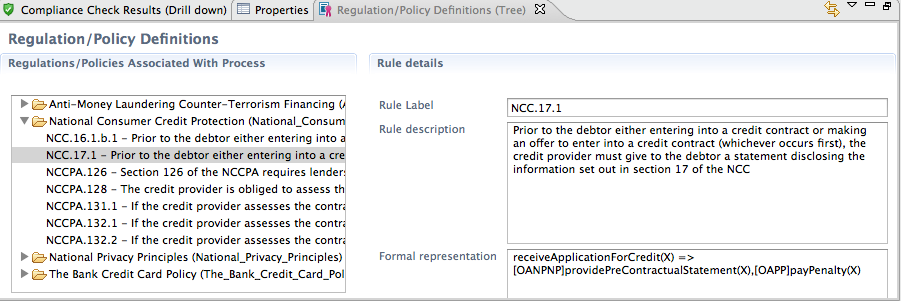}
  \caption{Regulations Relevant to the Opening Credit Card Process}
  \label{fig:label}
\end{figure*}


\section{Implementation and Evaluation} 
\label{sec:evaluation}

\begin{figure*}[htbp]
  \centering
    \includegraphics[width=\textwidth]{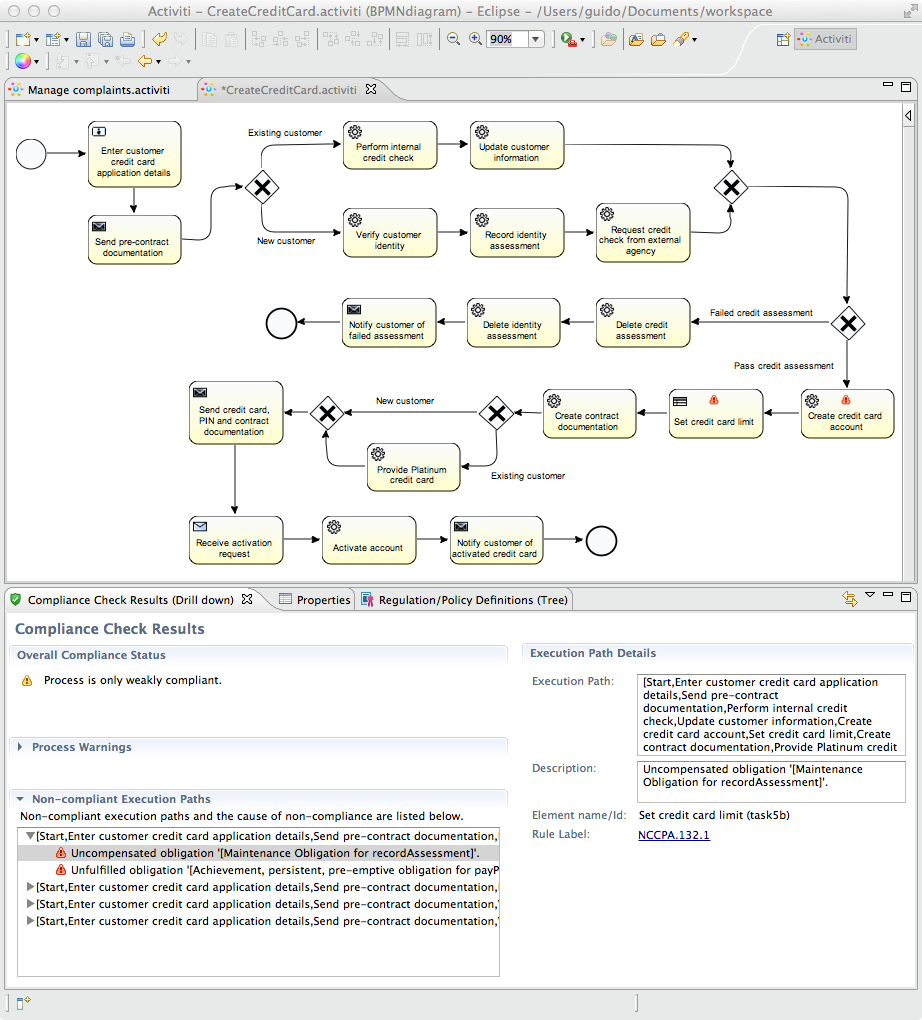}
  \caption{Regorous report of traces, rules, and tasks responsible for
  non-compliance}
  \label{fig:Figures_bpcc-compliance}
\end{figure*}

Regorous Process Designer is implemented on top of Eclipse. For the
representation of process
models, it uses the Eclipse Activiti BPMN 2.0 plugin, extended with features
to allow users to add semantic annotations to the tasks in the process model. 
Regorous is process model agnostic, this means that while the current
implementation is based on BPMN all Regorous needs is to have a description of the
process and the annotations for each task. A module of Regorous takes the
description of the process and generates the execution traces corresponding to
the process. After the traces are generated, it implements the algorithm
outlined in the previous section, where it uses the SPINdle rule engine
\cite{spindle} for the evaluation of the FCL rules. In case a process is not
compliant (or if it is only weakly compliant) Regorous reports the traces,
tasks, rules and obligations involved in the non compliance issues (see
Figure~\ref{fig:Figures_bpcc-compliance}).

Regorous was tested against the 2012 Australian Telecommunications Customers
Protection Code (C628-2012). The code is effective from September 1st 2012.
The code requires telecommunication operators to provide an annual attestation of
compliance with the code staring from April 1st 2013. The evaluation was
carried out in May-June 2012.
Specifically, the section of the code on complaint handling has been
manually mapped to FCL. The section of the code contains approximately 100
commas, in addition to approximately 120 terms given in the Definitions and
Interpretation section of the code. The mapping resulted in 176 FCL rules,
containing 223 FCL (atomic) propositions, and 7 instances of the superiority
relation. Of the 176 rules 33 were used to capture definitions of terms used
in the remaining rules. Mapping the section of the code required all features
of FCL. Table~\ref{tab:tcpc} reports the types of deontic effects present in 
the FCL mapping, and for each type the table includes the number of distinct 
occurrences and, in parenthesis, the total number of instances (some effects 
can have different conditions under which they are effective).

\begin{table}[htb]
\caption{Number and types of obligations and permissions in Section 8 of TCPC}
  \begin{center}
  \begin{tabular}{lrr}
  \toprule
  Punctual Obligation  & 5 & (5)\\
  \toprule
  Achievement Obligation & \quad 90 & (110)\\
  \midrule
  \multicolumn{1}{r}{Preemptive} & 41 & (46)\\
  \multicolumn{1}{r}{Non preemptive} & 49 & (64)\\
  \multicolumn{1}{r}{Non perdurant}   & 5  &  (7)\\
  \toprule
  Maintenance Obligation \qquad \qquad \qquad & 11 & (13)\\
  \midrule
   \multicolumn{1}{r}{Prohibition}  & 7  & (9)\\
   \multicolumn{1}{r}{Non perdurant} & 1  & (4)\\
  \toprule
  Permission    & 9  & (16)\\
  \toprule
  Compensation  & 2  & (2) \\
  \bottomrule
\end{tabular}
\end{center}
\label{tab:tcpc}
\end{table}

The evaluation was carried over in cooperation with an industry partner
subject to the code. The industry partner did not have
formalised business processes. Thus, we worked with domain experts from the
industry partner (who had not been previously exposed to BPM technology, but
who were familiar with the industry code) to draw process models for the
activities covered by the code. The evaluation was carried out in two steps.
In the first part we modelled the processes as they were. Regorous was able to
identify several areas where the existing processes were not compliant with
the new code. In some cases the industry partner was already aware of some of
the areas requiring modifications of the existing processes. However, some of
the compliance issues discovered by the tools were novel to the business
analysts and were identified as genuine non-compliance issues that need to be
resolved. In the second part of the experiment, the existing processes were
modified to comply with the code based on the issues identified in the first
phase. In addition a few new business process models required by the new code
were designed. As result we generated and annotated 6 process models. 5 of the
6 models are limited in size and they can be checked for compliance in
seconds. The largest process contains 41 tasks, 12 decision points, xor
splits, (11 binary, 1 ternary). The shortest path in the model has 6 tasks,
while the longest path consists of 33 tasks (with 2 loops), and the longest
path without loop is 22 task long. The time taken to verify compliance for
this process amounts approximately to 40 seconds on a MacBook Pro 2.2Ghz Intel
Core i7 processor with 8GB of RAM (limited to 4GB in Eclipse).


\section{Compliance at Design Time, Run Time and Auditing}
\label{sec:extensions}

The methodology and tool presented in the previous sections are primarily meant
to help in the design of compliant business processes according the principle
of compliance-by-design. While Regorous is implemented in a computer system
the proposed approach does not require the processes to be implement and
executed by a workflow engine. Obviously, an enterprise obtains major
benefits when the tasks in a process are fully automated and the coordination
of the order of execution of the task is under the control of a process-aware
information system (see \cite{DBLP:journals/is/DumasRW12} for an overview of
what process-aware information systems are and their functionalities). In such
a case, assuming a faithful implementation of the process, all instances of
the process are guaranteed to be compliant removing, potentially, the need of
run-time monitoring and post-execution auditing.

At the other extreme of the spectrum we have the case where processes are not
implemented by workflow engines. The proposed approach is still useful in so
far as it can be used to establish the blue-prints of compliant processes.
Clearly, if the tasks are executed by human operators (and the operators have
flexibility about what operations are executed, and when to execute them), the
tool cannot be used to support run-time monitoring and auditing, and other
well establish methods have to be used.

The last situation to consider is when there are no well defined process
models, but the business activities (i.e., processes) are still supported by
ICT technology in the form of recording of business events and message
passing, and writing them in a log. In this scenario, the approach we proposed
can be still applied. As we have outlined in
Section~\ref{sec:compliance} Regorous simulates all the possible (finite)
executions of a process, where an execution or trace is the sequence of tasks
to be executed. In this case we can use a business event as a task. Here,
instead of annotating the tasks in a process, we do the same on the business
events and messages to recorded (or to be recorded) in the log, and extract the data using the
techniques presented in \cite{ruleml2012}. At run-time, after each business
events Regorous can compute what are the obligations, prohibitions in force
after the business event, and evaluate whether they have been fulfilled or
violated and report the resulting state. For auditing, Regorous can examine
the log, and for each instance, replay it to determine, using the same
algorithms for compliance, whether the instance was properly executed, and if
it was compliant.

\section{Conclusions} 
\label{sec:conclusions}

We reported on the development of a tool, Regorous Process Designer, for checking the compliance of
business processes with relevant regulations. Regorous was successfully tested
for real industry scale compliance problems. In the recent years techniques and 
methodologies to address the problem of regulatory compliance from an ICT point 
of view have been proposed (see \cite{becker2012} for an extensive list of such 
approaches). Besides Regorous a few other
compliance prototypes have been proposed. Here we consider some representative ones: 
MoBuCom \cite{mobucom}, Compass \cite{compass} and
SeaFlows \cite{seaflows}. MoBuCom and Compass are based on Linear Temporal Logic (LTL)
and mostly they just address ``structural compliance'' (i.e., that the tasks
are executed in the relative order defined by a constraint model). The use of
LTL implies that the model on which these tools are based on is not conceptual
relative to the legal domain, and it fails to capture nuances of reasoning
with normative constrains such as violations, different types of obligations,
violations and their compensation. For example, obligations are represented by
temporal operators. This raises the problem of how to represent the
distinction between achievement and maintenance obligations. A possible
solution is to use always for maintenance and sometimes for achievement, but
this leaves no room for the concept of permission (the permission is dual of
obligation, and always and sometimes are the dual of each other). In addition
using temporal operators to model obligations makes hard to capture data
compliance \cite{ruleml2012}, i.e., obligations that refer to literals in the
same task. SeaFlow is based on first-order logic, and it is well know that
first oder logic is not suitable to capture normative reasoning
\cite{herrestad}. On the other hand FCL complies with the guidelines set up in
\cite{ruleml2009} for a rule languages for the representation of legal knowledge
and legal reasoning.

\section*{Acknowledgment} 
NICTA is funded by the Australian Government as represented by the Department
of Communication and the Australian
Research Council through the ICT Centre of Excellence program.

\printbibliography
\end{document}